\documentstyle[aps,prl,epsf]{revtex}

\begin{document}
\draft

\title{
Josephson current carried by Andreev levels \\
in superconducting quantum point contacts
}

\author{
Akira Furusaki}

\address{
Yukawa Institute for Theoretical Physics, Kyoto University,
Kyoto 606-8502, Japan$^*$\\
and Department of Physics, Stanford University, Stanford, California
94305-4060 
}

\date{\today}
\maketitle

\begin{abstract}
The dc Josephson effect in a superconducting
quantum point contact, where supercurrent flows through a small number
of channels, is reviewed.
The central role of Andreev levels is emphasized which carry the whole
supercurrent in short symmetric Josephson junctions including tunnel
junctions. 
A simple intuitive view of the dc Josephson effect in a quantum point
contact is given in terms of multiple Andreev reflections.
The quantization of the critical current in superconducting quantum
point contacts is briefly discussed.
\end{abstract}

\section{Introduction}
Recent surprising progress in microfabrication techniques has enabled
us to make tiny Josephson junctions with artificial geometry.
This opens a door to explore various mesoscopic effects in the
transport properties of small superconducting junctions.
In this paper we review the dc Josephson effect in
a superconducting quantum point
contact\cite{rf:PRL,rf:Beenakker-vanHouten},  
a Josephson junction consisting of two superconductors coupled through
a few quantum channels (Fig.~1).
Such a superconducting quantum point contact can be created in a
superconductor--2DEG--superconductor junction (2DEG: two-dimensional
electron gas) \cite{rf:Takayanagi} as well as in a mechanically
controllable break junction \cite{rf:Muller}.
Main focus will be on current-carrying bound states that appear near
a quantum point contact.
Our discussion is based on the Bogoliubov-de Gennes equation, which is
a two-component Schr\"odinger equation describing electron-like
and hole-like quasiparticle excitations in superconducting systems.
We show that the Bogoliubov-de Gennes equation has solutions
describing bound states that are localized around the junction and
decay exponentially into the superconductors.
These bound states play a central role in the dc Josephson effect.
The physical picture that emerges out of the solution is that the
supercurrent flows as a result of multiple Andreev reflections
\cite{rf:Andreev} that transform an electron into a hole or vice versa.
The bound states represent standing waves of electron and hole waves
going back and forth between two superconductors.
This view might look slightly different from a traditional picture of
the dc Josephson effect explained in textbooks, according to which the
supercurrent, calculated in lowest order in the tunneling Hamiltonian,
flows as a result of tunneling of Cooper pairs.
In fact, these views are simply two sides of a same coin.

\begin{figure}
\centerline{\epsfysize=1.5in\epsfbox{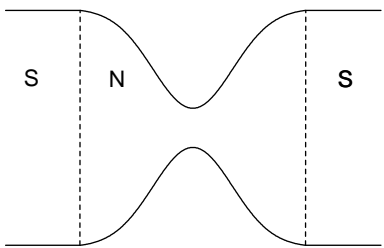}}
\caption{Superconducting quantum point contact.
The supercurrent flows through a narrow constriction in the normal
region. 
}
\label{fig:SQPC}
\end{figure}

The importance of the bound states caused by the multiple Andreev
reflections was first recognized in the early 1970s in the studies of
the dc Josephson effect in long superconductor--normal
metal--superconductor junctions
\cite{rf:Kulik,rf:Ishii,rf:Bardeen,rf:Svidzinsky}.
Until the 1990s, however, it was not fully appreciated that the
presence of current-carrying bound states is a general feature in any
symmetric Josephson junctions including weak links, superconducting
quantum point contacts, and conventional tunnel junctions.
The bound states in a short junction or a point contact, whose energy
depends on the phase difference of the superconductors, were
explicitly constructed by solving the Bogoliubov-de Gennes equation
in \cite{rf:Physica1,rf:Beenakker-vanHouten}.
It is a curious fact that almost no attention was paid to the
bound states in earlier studies of short superconducting weak links
\cite{rf:KO,rf:Zaitsev,rf:Arnold}, in which the bound states show up
silently as poles of the Green's function on the real axis.
In fact, the Andreev bound states play a central role for various
situations in mesoscopic superconducting systems.

In this paper we will concentrate on the dc Josephson effect of
$s$-wave superconductors.
When a finite voltage is applied across the junction, the loop of the
multiple Andreev scattering process is no longer closed, and
the solutions of the Bogoliubov-de Gennes equation contain
infinitely many terms each representing Andreev-reflected electron
wave or hole wave
\cite{rf:KTB,rf:Arte,rf:Flens,rf:Kummel,rf:Gunsen,rf:Bratus,rf:Averin,rf:Hurd}.
For sufficiently small applied voltage, however, the
current-voltage characteristics can be described in terms of the
quasi-stationary bound states whose energy is changing slowly with
time \cite{rf:Averin}.
We will not discuss further the ac Josephson effect and the
current-voltage characteristics of superconducting quantum point
contacts in this paper.

This paper is organized as follows.
In Sec.~II we introduce our basic equation, the Bogoliubov-de Gennes
equation, and a continuity equation for electric charge.
We solve a simplified model of a Josephson junction in Sec.~III to
illustrate the basic idea of the supercurrent carried by Andreev bound
states.
We point out that a midgap surface state in a $d$-wave superconductor
is essentially the same as the bound state in a short Josephson
junction. 
We then explain a scattering approach to calculate the dc Josephson
current in Sec.~IV.

\section{Bogoliubov-de Gennes equation}
We begin with the Bogoliubov-de Gennes equation\cite{rf:deGennes}
\begin{equation}
\pmatrix{
-\frac{\hbar^2}{2m}\frac{\partial^2}{\partial\mbox{\boldmath$r$}^2}
+U(\mbox{\boldmath$r$})-\mu &
 \Delta(\mbox{\boldmath$r$})\cr
\Delta^*(\mbox{\boldmath$r$}) &
 \frac{\hbar^2}{2m}\frac{\partial^2}{\partial\mbox{\boldmath$r$}^2}
 -U(\mbox{\boldmath$r$})+\mu \cr}
\pmatrix{u_n(\mbox{\boldmath$r$}) \cr v_n(\mbox{\boldmath$r$})}
=E_n\pmatrix{u_n(\mbox{\boldmath$r$}) \cr v_n(\mbox{\boldmath$r$})}.
\label{eq:BdG}
\end{equation}
where $U(\mbox{\boldmath$r$})$ represents both confining potential and
scattering potential at interfaces of a superconductor and a normal
metal, and $\mu$ is the chemical potential.
This is the basic equation for mesoscopic superconducting systems.
The upper component $u(\mbox{\boldmath$r$})$ corresponds to electrons
and the lower one $v(\mbox{\boldmath$r$})$ represents holes.
In principle the pair potential $\Delta(\mbox{\boldmath$r$})$ is
determined from the selfconsistency condition or the gap equation, 
$\Delta(\mbox{\boldmath$r$})=
\lambda\langle\Psi_\uparrow(\mbox{\boldmath$r$})
\Psi_\downarrow(\mbox{\boldmath$r$})\rangle$,
where $\lambda(>0)$ is a coupling constant of electron-electron
attraction, and $\langle\ \rangle$ is an average in the equilibrium
state. 
The field operator $\Psi_\sigma$ of electrons with spin $\sigma$ can
be expanded with the solutions of the Bogoliubov-de Gennes equation:
\begin{equation}
\pmatrix{\Psi_\uparrow(\mbox{\boldmath$r$},t)\cr
         \Psi^\dagger_\downarrow(\mbox{\boldmath$r$},t)}
=\sum_{E_n>0}\left[
a_{n\uparrow}e^{-iE_nt}\pmatrix{u_n(\mbox{\boldmath$r$})\cr
v_n(\mbox{\boldmath$r$})}
+a^\dagger_{n\downarrow}e^{iE_nt}
 \pmatrix{-v^*_n(\mbox{\boldmath$r$})\cr u_n^*(\mbox{\boldmath$r$})}
\right],
\label{eq:expansion}
\end{equation}
where the creation and annihilation operators $a_{n\sigma}$ and
$a^\dagger_{n\sigma}$ satisfy
$\{a_{n\sigma},a^\dagger_{n'\sigma'}\}
 =\delta_{n,n'}\delta_{\sigma,\sigma'}$
and $\{a_{n\sigma},a_{n'\sigma'}\}=0$.
Note that $(-v^*_n\ u^*_n)^{\rm t}$ is an eigenfunction with energy
$-E_n$ if $(u_n\ v_n)^{\rm t}$ satisfies Eq.\ (\ref{eq:BdG}).
That is why we sum over the positive energy states only and introduce
the second term in Eq.\ (\ref{eq:expansion}).

Following Ref.~\cite{rf:BTK}, we introduce the equation of continuity
of electric charge, which are satisfied by the solutions of the
Bogoliubov-de Gennes equation.
We define
\begin{mathletters}
\begin{eqnarray}
P_e&=&
-e\left[
\Psi^\dagger_\uparrow(\mbox{\boldmath$r$},t)
 \Psi_\uparrow(\mbox{\boldmath$r$},t)
+\Psi^\dagger_\downarrow(\mbox{\boldmath$r$},t)
 \Psi_\downarrow(\mbox{\boldmath$r$},t)
\right],
\label{eq:P_e}\\
\mbox{\boldmath$J$}_e&=&
-\frac{e\hbar}{m}{\rm Im}\left[
\Psi^\dagger_\uparrow(\mbox{\boldmath$r$},t)
\frac{\partial}{\partial\mbox{\boldmath$r$}}
 \Psi_\uparrow(\mbox{\boldmath$r$},t)
+\Psi^\dagger_\downarrow(\mbox{\boldmath$r$},t)
\frac{\partial}{\partial\mbox{\boldmath$r$}}
 \Psi_\downarrow(\mbox{\boldmath$r$},t)
\right].
\label{eq:J_e}
\end{eqnarray}
\end{mathletters}\noindent
It is clear that $P_e$ and $J_e$ are electric charge density and
electric current, respectively.
The continuity equation for the electric charge reads
$\partial_tP_e+{\rm div}\mbox{\boldmath$J$}_e+S=0$,
where the source term $S$ is given by
$S=
(4e/\hbar){\rm Im}
[
\Delta(\mbox{\boldmath$r$})
 \Psi^\dagger_\uparrow(\mbox{\boldmath$r$},t)
 \Psi^\dagger_\downarrow(\mbox{\boldmath$r$},t)
]$.
When the gap equation is satisfied, $\langle S\rangle$ vanishes, and
we recover the ordinary continuity equation of the charge for the
averaged quantities $\langle P_e\rangle$ and
$\langle J_e\rangle$, where $\langle\ \rangle$ is an average in
equilibrium. 
When we calculate the Josephson current in a model where the pair
potential satisfies the gap equation only approximately,
we have to take into account the contribution from $S$ \cite{rf:SSC}. 
The source term can be ignored when the current is calculated
inside the normal region where $\Delta(\mbox{\boldmath$r$})=0$.

\section{Current-carrying bound states}

We shall consider a simplest model of a Josephson junction that shows
the essential features of the Josephson effect.
It is a one-dimensional model where the left and right superconductors
have the pair potentials of the same magnitude but with different phases:
\begin{equation}
\Delta(x)=\cases{
\Delta_0e^{i\theta_L} &$x<0$,\cr
\Delta_0e^{i\theta_R} &$x>0$.\cr}
\label{eq:Delta(x)}
\end{equation}
We assume the normal region to be thin and consider an
extreme case where its width is infinitesimal.
In general we expect some scattering process to be present either
inside the normal region or at the super--normal interfaces.
We therefore introduce a scattering potential
\begin{equation}
U(x)=V\delta(x)\qquad V\ge0
\label{eq:U(x)}
\end{equation}
to describe such a scattering.
The model with (\ref{eq:Delta(x)}) and (\ref{eq:U(x)}) is the
simplest non-trivial model for the Josephson effect.
The strength of the $\delta$-function potential is characterized by a
dimensionless parameter $Z\equiv mV/\hbar^2k_F$, where $k_F$ is the
Fermi wave vector.
When $Z\ll1$ the model describes a short superconductor--normal
metal--superconductor junction.
In the opposite limit $Z\gg1$, the model can be regarded as a toy
model of a superconductor--insulator--superconductor tunnel junction.

We are interested in a state in the energy gap, $0<E<\Delta_0$.
In this energy range a wave function has to decay exponentially for
$|x|\to\infty$.
We can write the wave function as
\begin{equation}
\psi_B(x)=\pmatrix{u_B(x)\cr v_B(x)\cr}=\cases{
a_Be^{ik_hx}\pmatrix{v_0e^{i\theta_L/2}\cr u_0e^{-i\theta_L/2}\cr}
+b_Be^{-ik_ex}\pmatrix{u_0e^{i\theta_L/2}\cr v_0e^{-i\theta_L/2}\cr}
&$x<0,$\cr&\cr
c_Be^{ik_ex}\pmatrix{u_0e^{i\theta_R/2}\cr v_0e^{-i\theta_R/2}\cr}
+d_Be^{-ik_hx}\pmatrix{v_0e^{i\theta_R/2}\cr u_0e^{-i\theta_R/2}\cr}
&$x>0,$\cr}
\label{eq:psi_B}
\end{equation}
where
\[
k_e=\sqrt{\frac{2m}{\hbar^2}(\mu+i\eta)}\ ,\quad
k_h=k_e^*\ ,\quad
u_0=\sqrt{\frac{1}{2}\left(1+i\frac{\eta}{E}\right)}\ ,\quad
v_0=u_0^*\ ,\quad
\eta=\sqrt{\Delta^2_0-E^2}\ .
\]
The coefficients $a_B$, $b_B$, $c_B$, and $d_B$ are determined
from the matching condition at $x=0$:
\begin{equation}
\psi_B(-0)=\psi_B(+0),\qquad
\psi'_B(+0)-\psi'_B(-0)=\frac{2mV}{\hbar^2}\psi_B(0).
\label{eq:matching}
\end{equation}
The eigenenergy is determined from the condition that
Eq.\ (\ref{eq:matching}) has a nontrivial solution.
Within the approximation $k_e\approx k_h\approx k_F$ we
find\cite{rf:Physica1}
\begin{equation}
E=E_B\equiv\Delta_0\sqrt{\frac{\cos^2(\varphi/2)+Z^2}{1+Z^2}}.
\label{eq:E_B}
\end{equation}
The solution of Eq.\ (\ref{eq:matching}) with this energy is
\begin{eqnarray*}
a_B&=&
-N\sigma\sqrt{1+Z^2}\left[\sqrt{\cos^2(\varphi/2)+Z^2}
                           -\sigma\cos(\varphi/2)\right],\quad
b_B=
i\sigma NZ(1-iZ),\cr
c_B&=&
N(1-iZ)\left[\sqrt{\cos^2(\varphi/2)+Z^2}
              -\sigma\cos(\varphi/2)\right],\quad
d_B=
iNZ\sqrt{1+Z^2},
\end{eqnarray*}
where $\varphi=\theta_R-\theta_L$, $N$ is the
normalization constant 
\[
N^2=\frac{\sigma\Delta_0\sin(\varphi/2)}
       {2\hbar v_F(1+Z^2)^2
        \left[\sqrt{\cos^2(\varphi/2)+Z^2}
               -\sigma\cos(\varphi/2)\right]},
\]
and $\sigma={\rm sgn}(\varphi)$ for $-\pi\le\varphi\le\pi$.
The wave function is normalized such that
$
\int^\infty_{-\infty}\left[|u_B(x)|^2+|v_B(x)|^2\right]dx=1.
$
Once the wave function of the positive energy state is known, the
eigenstate with energy $-E_B$ is obtained as
$
\tilde\psi_B(x)=(-v^*_B(x),u^*_B(x))^t.
$
The field operator is then expanded as
\begin{equation}
\pmatrix{\Psi_\uparrow(x,t)\cr \Psi^\dagger_\downarrow(x,t)\cr}
=
a_{B\uparrow}e^{-iE_Bt}\psi_B(x)
+a^\dagger_{B\downarrow}e^{iE_Bt}\tilde\psi_B(x)
+ \,({\rm contribution~from~states~with}~|E|>\Delta_0).
\label{eq:expansion-B}
\end{equation}
Let us assume for a while that the states above or below the energy
gap ($|E|>\Delta_0$) do not contribute to the supercurrent.
We will see later that this is indeed the case for the model we are
dealing with.
We can calculate the supercurrent as the average of $J_e$ at $x=+0$ in
the equilibrium state at temperature $T$:
\begin{eqnarray}
I&=&\langle J_e\rangle
=-4ev_Fu_0v_0(|c_B|^2-|d_B|^2)[f(E_B)-f(-E_B)]\cr
&&\cr
&=&
-\frac{e\Delta_0\sin\varphi}{2\hbar(1+Z^2)}
\sqrt{\frac{1+Z^2}{\cos^2(\varphi/2)+Z^2}}
\tanh\!\left(\frac{E_B}{2k_BT}\right),
\label{eq:I_B}
\end{eqnarray}
where $f(E)=[\exp(E/k_BT)+1]^{-1}$.
Since the transmission probability through the $\delta$-function
barrier Eq.\ (\ref{eq:U(x)}) is ${\cal T}=1/(1+Z^2)$,
the normal-state conductance of this model is
$G_N=2e^2/[h(1+Z^2)]$.
We can thus rewrite Eqs.\ (\ref{eq:E_B}) and (\ref{eq:I_B}) as
\begin{mathletters}
\begin{eqnarray}
E_B&=&
\Delta_0\sqrt{1-{\cal T}\sin^2(\varphi/2)}\, ,
\label{eq:E_B-2}\\
I&=&
-G_N\frac{\pi\Delta_0}{2e}
 \frac{\sin\varphi}{\sqrt{1-{\cal T}\sin^2(\varphi/2)}}
 \tanh\!\left(\frac{E_B}{2k_BT}\right).
\label{eq:I_B-2}
\end{eqnarray}
\end{mathletters}\noindent
When ${\cal T}\ll1$ Eq.\ (\ref{eq:I_B-2}) reduces to
\begin{equation}
I=
-G_N\frac{\pi\Delta_0}{2e}\sin(\varphi)
 \tanh\!\left(\frac{\Delta_0}{2k_BT}\right),
\label{eq:AB}
\end{equation}
which is the well-known Ambegaokar-Baratoff
formula \cite{rf:Ambegaokar}.
In the opposite limit ${\cal T}=1$, Eq.\ (\ref{eq:I_B-2}) becomes
\begin{equation}
I=
-G_N\frac{\pi\Delta_0}{e}\sin(\varphi/2)
 \tanh\!\left(\frac{\Delta_0}{2k_BT}\cos(\varphi/2)\right),
\label{eq:KO}
\end{equation}
which coincides with the result obtained by Kulik and
Omel'yanchuk\cite{rf:KO} for a point contact Josephson junction.
In this limit the zero-temperature Josephson current becomes maximum
at $\varphi=\pi$, not at $\varphi=\pi/2$ like in the tunnel junctions.
The bound state energy is simply given by
$E_B=\Delta_0\cos(\varphi/2)$, which equals the Fermi energy $E_B=0$
at $\varphi=\pi$.

Equation (\ref{eq:I_B-2}) thus interpolates between
Eq.\ (\ref{eq:AB}) of tunnel junctions with very small transmission
probability and Eq.\ (\ref{eq:KO}) of point contacts with perfect
transmission. 
The same expression of the supercurrent interpolating the two limits
has been found by many people
\cite{rf:Haberkorn,rf:Zaitsev,rf:Arnold,rf:Glazman}. 
The phase-dependent bound state energy (\ref{eq:E_B}) and its
implication to the Josephson effect in short junctions were emphasized
in \cite{rf:Physica1}. 
The results were then generalized to the multichannnel contacts in
\cite{rf:Bee2}. 

Equation (\ref{eq:E_B}) shows that there is always a bound state
(Andreev level) with $E_B\ge0$ unless the phase difference $\varphi$ is
zero. 
This is true even for the tunnel junctions in which ${\cal T}\ll1$ and
$\Delta_0-E_B\ll\Delta_0$.
In the above model two bound states ($E=\pm E_B$) appear in
the energy gap $|E|<\Delta_0$ but disappear when $\varphi=0$.
This is a general property of short $s$-wave junctions where
$\hbar v_F/L\gg\Delta_0$ with $L$ being the length of the normal region
($L=0$ in our model).
In the opposite regime where $\hbar v_F/L\ll\Delta_0$, many
($\sim\hbar v_F/\Delta_0L$) Andreev levels exist in the gap for any
$\varphi$ \cite{rf:Kulik,rf:Ishii,rf:Bardeen,rf:Svidzinsky}.
These Andreev levels are easily understood as discrete levels of
normal electrons confined in a quantum well of ${\em offdiagonal}$
pair potential.
They are similar to the discrete energy levels in a vortex in an
$s$-wave type II superconductor \cite{rf:Caroli}.
A nontrivial observation we made in the above calculation is that, when
$\varphi\ne0$, two Andreev levels survive even in the limit $L\to0$,
in sharp contrast with a case where a particle is confined in a
quantum well of diagonal potential barrier, in which no bound
state exists in the limit $L\to0$.

\begin{figure}
\centerline{\epsfysize=1.5in\epsfbox{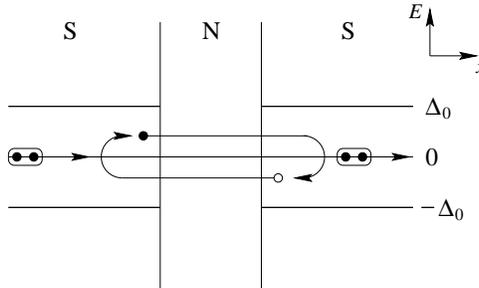}}
\caption{Current-carrying bound state.
The bound state is a manifestation of the multiple Andreev reflections
and carries the whole supercurrent in a short Josephson junction.
The filled (open) circle represent an electron (a hole).
The ovals containing two circles are Cooper pairs.}
\label{fig:andreev}
\end{figure}

The bound state $\psi_B(x)$ is formed as a result of resonant multiple
Andreev reflection process.
Equation (\ref{eq:I_B}) shows that this state carries nonvanishing
supercurrent.
This can be physically understood as follows (Fig.~\ref{fig:andreev}).
Each time an electron is Andreev reflected into a hole, a Cooper pair
is effectively generated.
Therefore the bound state, which represents an infinite loop of
Andreev reflections (electron$\to$hole$\to$electron$\cdots$), serves
as a pump that transfers Cooper pairs from one superconductor to the
other.
When ${\cal T}\approx 1$ the bound state energy has a strong
dependence on the phase difference $\varphi$, which can be controlled
externally by the current source.
It would be very interesting if these current-carrying Andreev levels
are observed experimentally using, for example, the scanning tunneling
spectroscopy.

\begin{figure}
\centerline{\epsfysize=2in\epsfbox{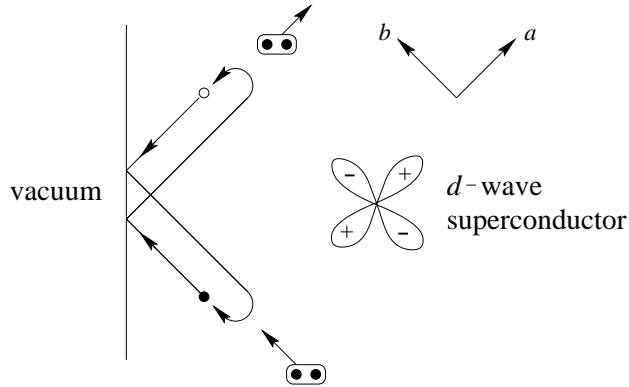}}
\caption{Midgap state formed near the surface of a $d$-wave
  superconductor.}
\label{fig:dwave}
\end{figure}

Finally we comment on midgap states that appear near the surface of a
$d$-wave superconductor when the surface is not parallel to a crystal
axis (Fig.~\ref{fig:dwave}) \cite{rf:Hu}.
In fact the mechanism of the formation of the zero-energy midgap state
is the same as the Andreev bound states we discussed above.
As shown in Fig.~\ref{fig:dwave}, an electron traveling towards a
surface is reflected back into the $d$-wave superconductor and is
subsequently Andreev-reflected into a hole by the {\it positive} pair
potential. 
In the next step the hole follows the same path backwards, reflected
at the surface, and finally Andreev-reflected into
another electron by the {\it negative} pair potential.
The analogy to the Josephson junction is now obvious.
The surface of the $d$-wave superconductor plays a role of the point
contact with ${\cal T}=1$, and the sign change in the pair potential
corresponds to the phase difference $\varphi=\pi$.
Equation (\ref{eq:E_B-2}) tells that there must be bound states with
$E_B=0$ localized near the surface, which are the midgap states found
in Ref.~\cite{rf:Hu}.
They give rise to a peak in differential conductance of a
normal--insulator--$d$-wave tunnel junction \cite{rf:Tanaka}.
The Andreev levels also play an important role in a Josephson
junction made of an $s$-wave superconductor and a $d$-wave
superconductor \cite{rf:Yip,rf:Zagoskin}. 
It is known that in such a junction there is a chance that Andreev
levels appear even in the lowest-energy state, thereby leading to a
ground state with broken time-reversal symmetry \cite{rf:Sigrist}.

\section{Josephson current from scattering amplitudes}
We shall introduce a general expression of the dc Josephson current
under the assumption that we can ignore the spatial variation of the
pair potential.
Here we assume
\begin{equation}
\Delta(\mbox{\boldmath$r$})=\cases{
\Delta_Le^{i\theta_L} &$x<0$,\cr
\Delta_Re^{i\theta_R} &$x>L$.\cr}
\label{eq:Delta(r)}
\end{equation}
In the normal region ($0<x<L$) the order parameter vanishes.
We also assume that the potential $U(\mbox{\boldmath$r$})$ is
independent of $x$ in the superconductors. 
We emphasize that this step-function pair potential
(\ref{eq:Delta(r)}) is a good approximation for mesoscopic Josephson 
junctions of our concern, where bulk superconductors are coupled via a
few conduction channels.
The use of the step-function model is justified by the following
reasons. 
First, the presence of a point contact does not seriously affect the
order parameter of bulk superconductors, because the width of the
constrictions is much smaller than the superconducting
coherence length $\xi$, which is the characteristic length scale for
the variation of the order parameter.
Second, if the normal region is made of a material different from
superconducting electrodes, the contact resistance of the SN interfaces
is not small due to the Fermi-velocity difference and/or due to the
presence of a Schottky barrier.
In this case also the pair potential is not reduced significantly near
the point contact.
More generally, when the resistance of the junction is much larger than 
those of superconductors, we may safely use the step-function
model \cite{rf:Likharev}.

Since the dc Josephson current is an equilibrium current, it is given
by the thermal average of $\mbox{\boldmath$J$}_e$.
We calculate it at $x=0$ to avoid complication due to the source term
$S$: 
\begin{equation}
I=\langle\mbox{\boldmath$e$}_x\cdot\mbox{\boldmath$J$}_e\rangle=
\frac{e\hbar}{2im}\lim_{x\to0}\lim_{x'\to x}\int dydz
\left(\frac{\partial}{\partial x}-\frac{\partial}{\partial x'}\right)
\frac{1}{\beta}\sum_{\omega_n}
 {\rm Tr}[G_{\omega_n}(\mbox{\boldmath$r$},\mbox{\boldmath$r$}')]
\biggr|_{x=0},
\label{eq:I=G_omega_n}
\end{equation}
where $G_{\omega_n}(\mbox{\boldmath$r$},\mbox{\boldmath$r$}')$ is the
temperature Green's function at frequency $\omega_n=\pi k_BT(2n+1)$ in
the Nambu space ($2\times2$ matrix) \cite{rf:AGD}.
Our assumption that the order parameter $\Delta(\mbox{\boldmath$r$})$ and
the potential $U(\mbox{\boldmath$r$})$ have no dependence on $x$ in the
superconductors, is strong enough that we can immediately write
down $G_{\omega_n}(\mbox{\boldmath$r$},\mbox{\boldmath$r$}')$  at
$x<0$ in terms of scattering amplitudes.
Its explicit expression can be found in Ref.~\cite{rf:SSC}.
After substituting it into Eq.\ (\ref{eq:I=G_omega_n}), we obtain
\begin{equation}
I=
\frac{e\Delta_L}{\hbar\beta}
\sum_{\omega_n}\frac{1}{\Omega_{nL}}
\sum_j[a_{jj}(\varphi,i\omega_n)-a_{jj}(-\varphi,i\omega_n)],
\label{eq:I=a-a}
\end{equation}
where $\Omega_{nL}=\sqrt{\omega^2_n+\Delta^2_L}$ and we have used the
Andreev approximation where the velocities of quasiparticles are set
equal to the Fermi velocity.

The quantity $a_{jk}(\varphi,i\omega_n)$ is a scattering amplitude for
the process in which an electron-like quasiparticle of $j$th channel
traveling from the left to the junction is reflected back as a
hole-like quasiparticle of $k$th channel \cite{rf:SSC}.
Equation (\ref{eq:I=a-a}) expresses the supercurrent in terms of the
scattering data $a_{jj}$, just like the Landauer formula does the
normal-state conductance. 
An important difference is that the supercurrent is proportional to
the scattering amplitude, not to the probability.
Equation (\ref{eq:I=a-a}) is clearly an odd function of $\varphi$, and
can be expanded formally as
$I=\sum_{m=1}^\infty I_m\sin(m\varphi)$.
The term proportional to $\sin(m\varphi)$ corresponds to $m$th order
term in the tunneling Hamiltonian approach that is a contribution from
the simultaneous tunneling of $m$ Cooper pairs.
Equation (\ref{eq:I=a-a}) is thus a useful, nonperturbative formula
for the dc Josephson current \cite{rf:Physica2}.

\begin{figure}
\centerline{\epsfysize=1.5in\epsfbox{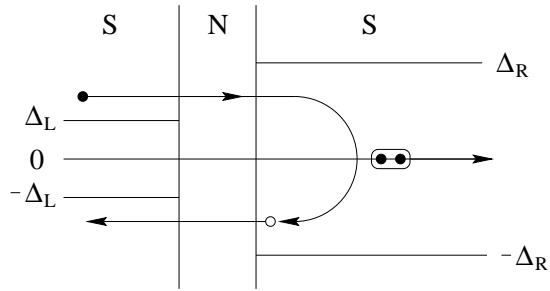}}
\caption{Scattering states carrying supercurrent in a short asymmetric
  junction.}
\label{fig:asymmetry}
\end{figure}

Let us apply it to the simple one-dimensional model we analyzed in the
last section.
Solving the Bogoliubov-de Gennes equation for $E>\Delta_0$ with
potentials given by (\ref{eq:Delta(x)}) and (\ref{eq:U(x)}) gives
\begin{equation}
a(\varphi,E)=
-\frac{\Delta_0[E\sin^2(\varphi/2)
        +i\Omega\sin(\varphi/2)\cos(\varphi/2)]}
      {E^2(1+Z^2)-\Delta^2_0[\cos^2(\varphi/2)+Z^2]}.
\label{eq:a(varphi,E)}
\end{equation}
Note that the scattering amplitude has poles at $E=\pm E_B$.
Substituting (\ref{eq:a(varphi,E)}) with $E=i\omega_n$ into
(\ref{eq:I=a-a}), we then find
\begin{equation}
I=
-\frac{e\Delta_0^2}{\hbar\beta}\sum_{\omega_n}
\frac{\sin\varphi}{\omega^2_n(1+Z^2)+\Delta^2_0[\cos^2(\varphi/2)+Z^2]},
\label{eq:I=sum_omegan}
\end{equation}
which agrees with Eq.~(\ref{eq:I_B}) after the frequency sum is
evaluated by reducing into a contour integral.
In this calculation all the contributions to the integral come from
the residues at $E=\pm E_B$.
This proves that the Andreev bound states carry the whole supercurrent
in short symmetric Josephson junctions.
We note that this is not true for long weak links where
$\hbar v_F/L\ll\Delta_0$ \cite{rf:Ishii,rf:Svidzinsky} and for
asymmetric junctions where $\Delta_L<\Delta_R$.
Let us examine the latter case in more detail.
We consider a one-dimensional model where the pair potential is given
by Eq.~(\ref{eq:Delta(r)}) with $L=0$ and the diagonal potential is
(\ref{eq:U(x)}).
A solution of the Bogoliubov-de Gennes equation for this model gives a
scattering amplitude for $E>\Delta_R>\Delta_L$, 
\begin{equation}
a(\varphi,E)=
\frac{E(\Delta_R\cos\varphi-\Delta_L)-i\Delta_R\Omega_L\sin\varphi}
     {E^2+K\Omega_R\Omega_L-\Delta_R\Delta_L\cos\varphi},
\label{eq:a(varphi)-asym}
\end{equation}
where $\Omega_{R(L)}=(E^2-\Delta^2_{R(L)})^{1/2}$ and $K=1+2Z^2$.
From Eq.~(\ref{eq:I=a-a}) we obtain
\begin{equation}
I=-\frac{2e}{\hbar\beta}\sum_{\omega_n}
\frac{\Delta_R\Delta_L\sin\varphi}
     {\omega^2_n+K\Omega_{nR}\Omega_{nL}+\Delta_R\Delta_L\cos\varphi},
\label{eq:I-asym}
\end{equation}
where $\Omega_{nR(L)}=(\omega^2_n+\Delta^2_{R(L)})^{1/2}$.
This result agrees with the Ambegaokar-Baratoff formula
\cite{rf:Ambegaokar} for a tunnel junction in the limit $K\gg1$ and
with the Kulik-Omel'yanchuk formula \cite{rf:KO} for a point contact
with perfect transmission in the limit $K\to1$.
The summation can be reduced to an integral over
$\Delta_L<|E|<\Delta_R$ plus contributions from residues at energies
of bound states at $|E|<\Delta_L$.
The integral over $\Delta<|E|<\Delta_R$ corresponds to the process
shown in Fig.~\ref{fig:asymmetry} and carry most of the supercurrent.

\section{Quantization of supercurrent}
In a superconducting quantum point contact, the critical current or
the maximum value of the dc Josephson current can take a quantized
value in units of $e\Delta_0/\hbar$.
This was first predicted theoretically
\cite{rf:PRL,rf:Beenakker-vanHouten} and 
confirmed by experiments \cite{rf:Takayanagi,rf:Muller}.
The effect can be understood easily by the following argument.
The critical current of a Josephson junction is generally proportional
to its normal-state conductance, as seen in Eq.\ (\ref{eq:I_B-2}).
It is well known that the conductance of a quantum point contact
fabricated in a 2DEG is quantized in units of $2e^2/h$
\cite{rf:vanWees,rf:Wharam}:
The conductance is simply $2e^2/h$ times the number of open channels
at the constriction.
We can thus naively expect that the critical current should be
quantized as well.
For a more quantitative discussion we can use the adiabatic
approximation where scattering between channels is neglected.
The transmission probability through the quantum point contact is then
${\cal T}_j=1$ for open channels and ${\cal T}_j=0$ for closed channels.
For a short symmetric superconducting quantum point contact where
$\hbar v_F/L\gg\Delta_0$, we obtain, from Eqs.\ (\ref{eq:I_B-2}) and
(\ref{eq:I=a-a}),
\begin{equation}
I=-N\frac{e\Delta_0}{\hbar}\sin(\varphi/2)
   \tanh\!\left(\frac{\Delta_0}{2k_BT}\cos(\varphi/2)\right),
\label{eq:NKO}
\end{equation}
where $N$ is the number of open channels at the point contact.
We thus find that the critical current is a multiple of
$e\Delta_0/\hbar$ at zero temperature.
The number of open channels is determined by the width of the
constriction, which can be controlled externally, for example,
by changing a gate voltage for a 2DEG.
This was done in an experiment by Takayanagi {\it et al.}
\cite{rf:Takayanagi}.

Equation (\ref{eq:NKO}) is valid only for short junctions.
In the opposite case where $\hbar v_F/L\ll\Delta_0$, the unit of the
quantization is different.
The supercurrent for a long superconducting quantum point contact is
given by
\begin{equation}
I\approx
-\frac{e}{\pi L}\sum^N_{j=1}v_j\varphi
\label{eq:NIshii}
\end{equation}
for $|\varphi|<\pi$ at zero temperature, where $v_j$ is the
longitudinal component of the Fermi velocity in $j$th channel.
In this case the critical current is not exactly quantized in contrast
with the short-junction limit.
This is nothing but a quantum limit of long SNS junctions obtained in
early 1970s \cite{rf:Ishii,rf:Bardeen}.

\section{Conclusions}
In this paper we have reviewed the dc Josephson effect in mesoscopic
Josephson junctions.
Our discussion is based on a simple analysis of the Bogoliubov-de
Gennes equation.
In short junctions including superconducting
quantum point contacts the whole supercurrent is carried by
Andreev bound states in the energy gap that appear as a result of
multiple Andreev reflections.
The wave function of a bound state is constructed explicitly for
a simple $\delta$-function model by solving the Bogoliubov-de Gennes
equation.
The bound state energies are sensitive to the difference of the
phases of superconductors, and this sensitivity is the origin of the
Josephson effect.
Our simple physical view of the Josephson effect in terms of the
current-carrying bound states is very general and applicable to many
situations.

\acknowledgements
The author thanks M.~Sigrist and A.~M.~Zagoskin for helpful comments
on the manuscript.

\end{document}